\documentclass[namedreferences]{solarphysics}
\usepackage[optionalrh,solaromanenum]{spr-sola-addons} % For Solar Physics
\usepackage{natbib}                     % For eps figures, old commands
\usepackage{graphicx}                    % For eps figures, newer & more powerfull
\usepackage{color}                       % For color text: \color command
\chardef\us=`\_

\begin{document}

\begin{article}

\begin{opening}

\title{The Grad-Shafranov Reconstruction of Toroidal Magnetic Flux Ropes:
First Applications}

%%%%%%%%%%%%%%%%%%%%%%%%%%%%%%%%%%%%%%%%%%%%%%%%%%%
%% Authors Names
%
 \author[addressref={aff1},email={qiang.hu.th@dartmouth.edu}]{\inits{Q.~}\fnm{Qiang}~\lnm{Hu}\orcid{0000-0002-7570-2301}}
% Mark Linton, B. Wood, P. Riley, Teresa Chinchilla
 \author[addressref={aff2},email={mark.linton@nrl.navy.mil}]{\inits{M.~G.}\fnm{Mark~G.}~\lnm{Linton}}%\orcid{0000-0002-7570-2301}}
\author[addressref={aff2},email={brian.wood@nrl.navy.mil}]{\inits{B.~E.}\fnm{Brian~E.}~\lnm{Wood}}
\author[addressref={aff3},email={pete@predsci.com}]{\inits{P.~}\fnm{Pete}~\lnm{Riley}}
\author[addressref={aff2,aff4},email={teresa.nieves-chinchil-1@nasa.gov}]{\inits{T.~}\fnm{Teresa}~\lnm{Nieves-Chinchilla}}

%%%%%%%%%%%%%%%%%%%%%%%%%%%%%%%%%%%%%%%%%%%%%%%%%%%
%% Runningheads
%
\runningauthor{HU et al.} \runningtitle{GS Reconstruction in
Toroidal Geometry}

%%%%%%%%%%%%%%%%%%%%%%%%%%%%%%%%%%%%%%%%%%%%%%%%%%%
%% Affilations
%% id shold be the same with \author addressref value.
\address[id={aff1}]{Department of Space Science and CSPAR, The University of Alabama in Huntsville, Huntsville, AL 35805}
\address[id={aff2}]{Naval Research Laboratory, Space Science Division, Washington, DC 20375}
\address[id={aff3}]{PSI, San Diego, CA}
\address[id={aff4}]{Catholic University of America, Washington,
DC}
%%%%%%%%%%%%%%%%%%%%%%%%%%%%%%%%%%%%%%%%%%%%%%%%%%%
%%% Abstract
\begin{abstract}
This article completes and extends a recent study of the
Grad-Shafranov (GS) reconstruction in toroidal geometry, as
applied to a two and a half dimensional configurations in space
plasmas with rotational symmetry. A further application to the
benchmark study of an analytic solution to the toroidal GS equation
with added noise shows deviations in the reconstructed geometry of
the flux rope configuration, characterized by the orientation of
the rotation axis, the major radius, and the impact parameter. On
the other hand, the physical properties of the flux rope, including
the axial field strength, and the toroidal and poloidal magnetic flux,
agree between the numerical and exact GS solutions. We also
present a real event study of a magnetic cloud flux rope from
\textit{in situ} spacecraft measurements. The devised procedures
for toroidal GS reconstruction are successfully executed. Various
geometrical and physical parameters are obtained with associated
uncertainty estimates. The overall configuration of the flux rope
 from the GS reconstruction is
compared with the corresponding morphological reconstruction based
on white-light images. The results show overall consistency,  but also
discrepancy in that the inclination angle of the flux rope central
axis with respect to the ecliptic plane differs by about 20-30
degrees in the plane of the sky. We also compare the results with
the original straight-cylinder GS reconstruction and discuss our
findings.
\end{abstract}

%%%%%%%%%%%%%%%%%%%%%%%%%%%%%%%%%%%%%%%%%%%%%%%%%%%
%% Keywords
%
%\keywords{}


\end{opening}
%-------------------------------------------------

%%%%%%%%%%%%%%%%%%%%%%%%%%%%%%%%%%%%%%%%%%%%%%%%%%%
%% Sections
%
 \section{Introduction}\label{s:intro}
This article is a continuation of the recently completed study,
now in press \citep[][hereafter Paper I]{2017Husolphys},  which
describes in detail the two-step recipe for the Grad-Shafranov
(GS) reconstruction of magnetic flux ropes, or magnetic clouds
(MCs), with a toroidal geometry. Under such a geometry, the local
configuration of the flux rope in the vicinity of the spacecraft
path is approximated by a section of a torus, instead of a
straight cylinder. But the symmetry of the structure remains,
changing from the translation symmetry of the former
straight-cylinder case to the rotational symmetry in the toroidal
geometry, so that the central axis of the flux rope in the latter
case possesses a finite radius (non-zero curvature).

Under these two-dimensional (2D) or 2.5D configurations with
non-vanishing axial magnetic field component, the Grad-Shafranov
(GS) equation applies for both configurations, governing the
magnetic and plasma structure in magnetohydrostatic equilibrium.
The GS reconstruction technique has been developed and applied to
reconstruct the 2D cross section of cylindrical magnetic and
plasma configuration, using \textit{in situ} spacecraft
measurements, for the past twenty years. The technique first
originated from the application to the magnetopause current sheet
crossings  \citep{1996GeoRLS,1999JGRH}. Later it was applied to
reconstructing cross sections of magnetic flux ropes, including
applications to small-scale flux ropes in the solar wind
\citep{2001GeoRLHu} and  to large-scale MCs \citep{2002JGRAHu} for
the first time. Additional applications include reconstruction of
flux transfer events (FTEs) in Earth's magnetopause, plasmoids in
the tail, and even recently to flux ropes in the Martian
atmosphere \citep{2014JGRA..119.1262H,2016GeoRL..43.4816H}. The
extensions of the GS reconstruction technique to GS-type
applications were also developed, which goes beyond static
equilibrium, especially allowing for description of dynamic
equilibrium involving significant remaining flow and inertial
forces, but remaining in a cylindrical geometry \citep[][and
references therein]{Hu2017GSreview}.  Recent extension to a
toroidal geometry was fully developed and reported in Paper I, with
detailed benchmark studies.

A recent review, commemorating the occasion of twenty year's
application of the GS reconstruction method, is presented by
\citet{Hu2017GSreview}. That article reviews the history of the
development of the technique and summarizes its main applications
to recovering various 2.5D space plasma structures, including the
magnetopause current sheet, the FTEs, flux ropes in geotail, and
magnetic flux ropes (and MCs) in the solar wind. In particular,
emphasis was put on validation of GS reconstruction results by
direct co-spatial multi-spacecraft measurements and indirect
quantitative correlation among physical properties between
\textit{in situ} and solar source region analysis, especially
through inter comparison of magnetic flux and helicity content for
MCs with associated flares and coronal mass ejections (CMEs)
\citep{Qiu2007,2012SoPhK,2014ApJH}. In addition, field-line twist
and length distributions inside MCs were also derived from the GS
reconstruction results and were used to assist in interpretation
of flux rope configuration and formation through magnetic
reconnection processes \citep{2015JGRAH,2014ApJH}. A qualitative
comparison of the GS reconstruction with toroidal geometry to the
numerical simulation result of \citet{2004JASTPR} was also
presented in \citet{Hu2017GSreview} as an extension to the
original GS reconstruction technique with straight-cylinder
geometry.

Traditionally, the modeling of magnetic flux rope configurations
based on the \textit{in situ} single spacecraft measurements is performed 
via least-squares fitting of a specific analytic model to the
time-series data (mostly magnetic field vectors).  The use of a
toroidal flux rope model was  developed by
\citet{1997GMS....99..147M} and \citet{2003GeoRL..30.2065R}, usually with a
circular cross section and a large aspect ratio (the ratio between
the major and minor radii of the torus). Hidalgo and colleagues
have also developed over the years a sophisticated algorithm for
fitting \textit{in situ} spacecraft measurements with a flux rope
model of non-circular cross section, a global configuration
conforming to a 3D rope structure rooted on the Sun, and/or
non-field aligned current \citep[\textit{e.g.}][and references
therein]{2016ApJ...823...27N,2016ApJ...823....3H}. The GS
reconstruction differs from these fitting based approaches by
yielding a completely 2D cross section as a numerical solution to
a more general non-force free state, although fitting is also
involved but on one magnetic field component only (the axial field
component).  Another difference is that the current implementation
of the toroidal GS reconstruction does not apply to the situation
of a submerged spacecraft path, \textit{i.e.} a path not exiting
into the center hole of the torus, as we illustrated in detail in
Paper I. In this situation, one has to resort to the traditional
fitting method for a toroidal geometry, as advocated by
\citet{2015SoPh..290.1371M}. Some interesting  perspectives of
\textit{in situ} flux rope modeling that go beyond static or 2D
configurations were also discussed in \citet{Hu2017GSreview},
especially regarding an intrinsically 3D configuration with
ribbon-like structures often present in solar prominence
observations, resembling a ``stellarator" configuration in fusion
devices \citep{freidberg87}.

The current article is organized as follows. In the next two sections, we present two case studies: one being a further benchmark study continuing from Paper I, and the other being a real event study observed by the Solar Terrestrial Relations Observatory (STEREO) B spacecraft, following the procedures delineated in Paper I. In particular, we present a comparison of the toroidal GS reconstruction result with the morphological modeling based on white-light images of the real event. Lastly we conclude and summarize our results.

\section{Application to a Benchmark Case}\label{sec:bench}

We have performed the toroidal GS reconstruction study of the
numerical simulation result of \citet{2004JASTPR}, where a flux rope configuration with exact toroidal geometry, but with  full MHD and temporal evolution, was generated in the computational domain, covering the inner heliosphere from the Sun to $\sim$1~AU. However a
quantitative comparison of various physical quantities cannot be
made, because the quantitative results of that simulation cannot
be recovered due to the long time period elapsed.  The qualitative
comparison in terms of the orientation, the basic geometry and
shape of the flux rope is presented and discussed in
\citet{Hu2017GSreview}. 

Here we present a benchmark study that is
one step further than the one presented in Paper I, where the
benchmark study of an analytic case  with  general toroidal configuration and
 noise addition terminated after the determination of the
rotation axis $Z$ and the major radius $R=r_0$, without the final
reconstruction step of applying the GS solver to obtain a solution
to the toroidal GS equation. In what follows, we complete this
last step and present a quantitative comparison between the exact
and numerical solutions, following the procedures given in Paper
I, as to be employed when examining a real event.

 \begin{figure}
 \centerline{\includegraphics[width=1.\textwidth,clip=]{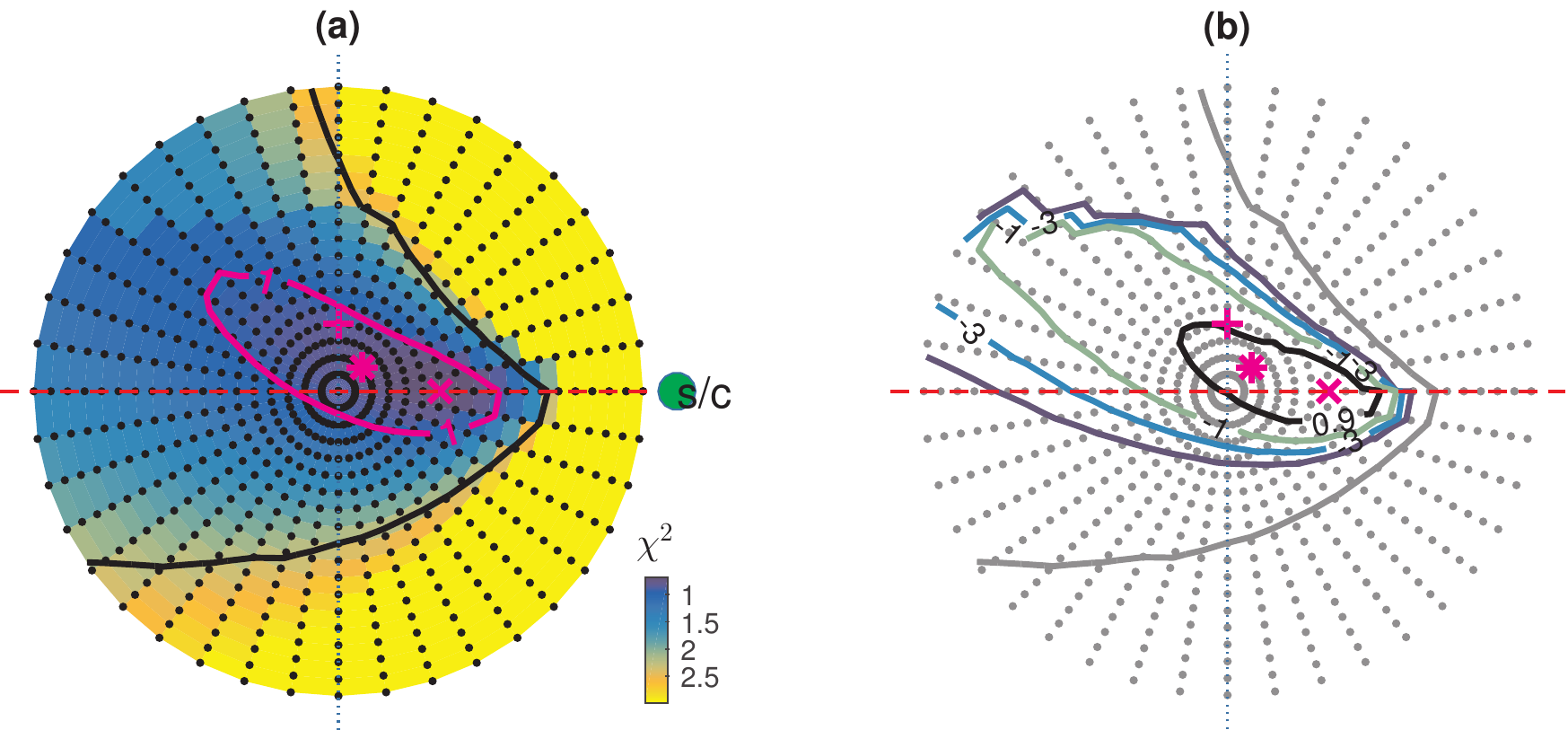}}
 \caption{(a) The distribution of reduced $\chi^2$ value, as indicated by the colorbar  on the $(r_{sc},t)$ plane within the circular domain centered on the Sun with a radius of 1~AU. The background
 dots are the search grid of the intersections of $Z$ axis on that plane. The contours are of levels 1 (magenta), and 1+$\sqrt{2}$ (black), respectively. (b) The corresponding contour plot of $\log_{10}Q$ as labeled.
 The innermost contour is of level $Q=0.9$, and the outermost one is the same as the black one in (a).   In each plot, the coordinate system $(r_{sc}, t, n)$ is centered on the Sun in the middle with $r_{sc}$ pointing radially out to the spacecraft location (green dot in a) along the red dashed line, $t$ pointing upward along the dotted blue line, and $n$ pointing vertically out of the plane at the center. The plus and cross signs
 mark the true $Z$ axis location and that of $\chi^2_{min}$, respectively. The asterisk  sign near the center of the $\chi^2=1$ contour denotes the one chosen for the final numerical GS reconstruction.}\label{fig:chi2}
 \end{figure}

We refer readers to Paper I for detailed description of the
two-step  recipe in deriving the main geometrical parameters and
the implementation of the GS solver for the toroidal GS equation,
using \textit{in situ} spacecraft data. In short, a trial and
error process is carried out to find the optimal orientation of
the rotation axis, $Z$, around which the rotational symmetry is
satisfied, \textit{i.e.} in the cylindrical $(R,\phi,Z)$
coordinate system, \textit{i.e.} $\partial/\partial\phi\approx 0$. The cross
sectional plane is $(R,Z)$. In the first step, the orientation of
$Z$ axis is determined by fulfilling the requirement that the
composite function $F=RB_\phi$ be a single-valued function of the
magnetic flux function $\Psi$ along the spacecraft path. Then, a
$\chi^2$ minimization procedure \citep{2002nrca.book.....P} is
performed by minimizing the deviation between the measured and GS
model predicted magnetic field components along the spacecraft
path, while altering the location of the $Z$ axis over a search
grid on the $(r_{sc}, t)$ plane, as illustrated in
Figure~\ref{fig:chi2}a. During this second step, a proper
formulation following \citet{2002nrca.book.....P} is adopted with
measurement uncertainties assigned to each data point. An optimal
location, or the intersection of $Z$ axis with the $(r_{sc},t)$
plane is chosen with uncertainty bounds, which in turn yields the
major radius $r_0$ of the torus. These uncertainty bounds lead to
estimates of uncertainty of other associated quantities as to be
discussed later. Once the main geometrical parameters, $Z$ and
$r_0$, are determined by the two-step procedures, the final
reconstruction is performed by applying the toroidal GS solver
 to obtain the cross section of the flux rope in an
annular region (Paper I).

\begin{figure}
 \centerline{\includegraphics[width=0.5\textwidth,clip=]{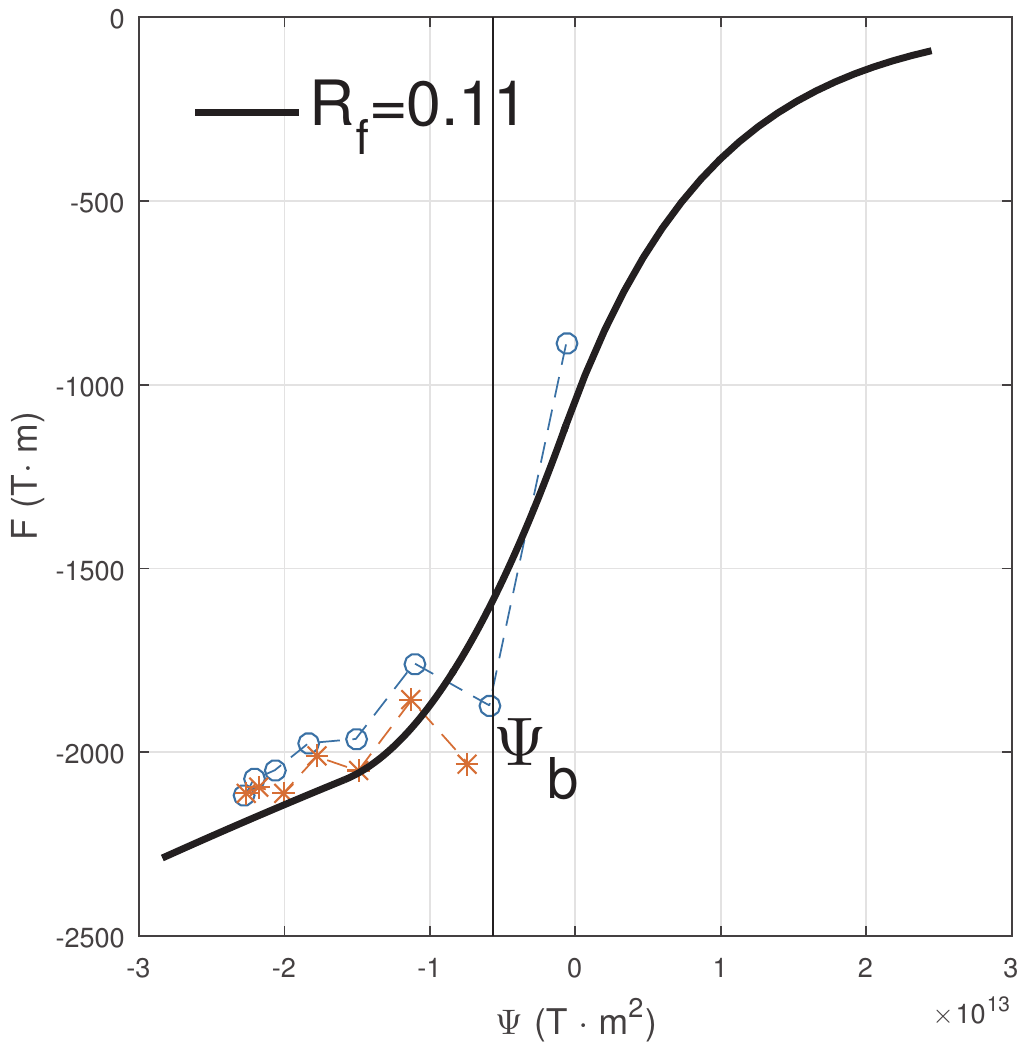}
 \includegraphics[width=0.5\textwidth,clip=]{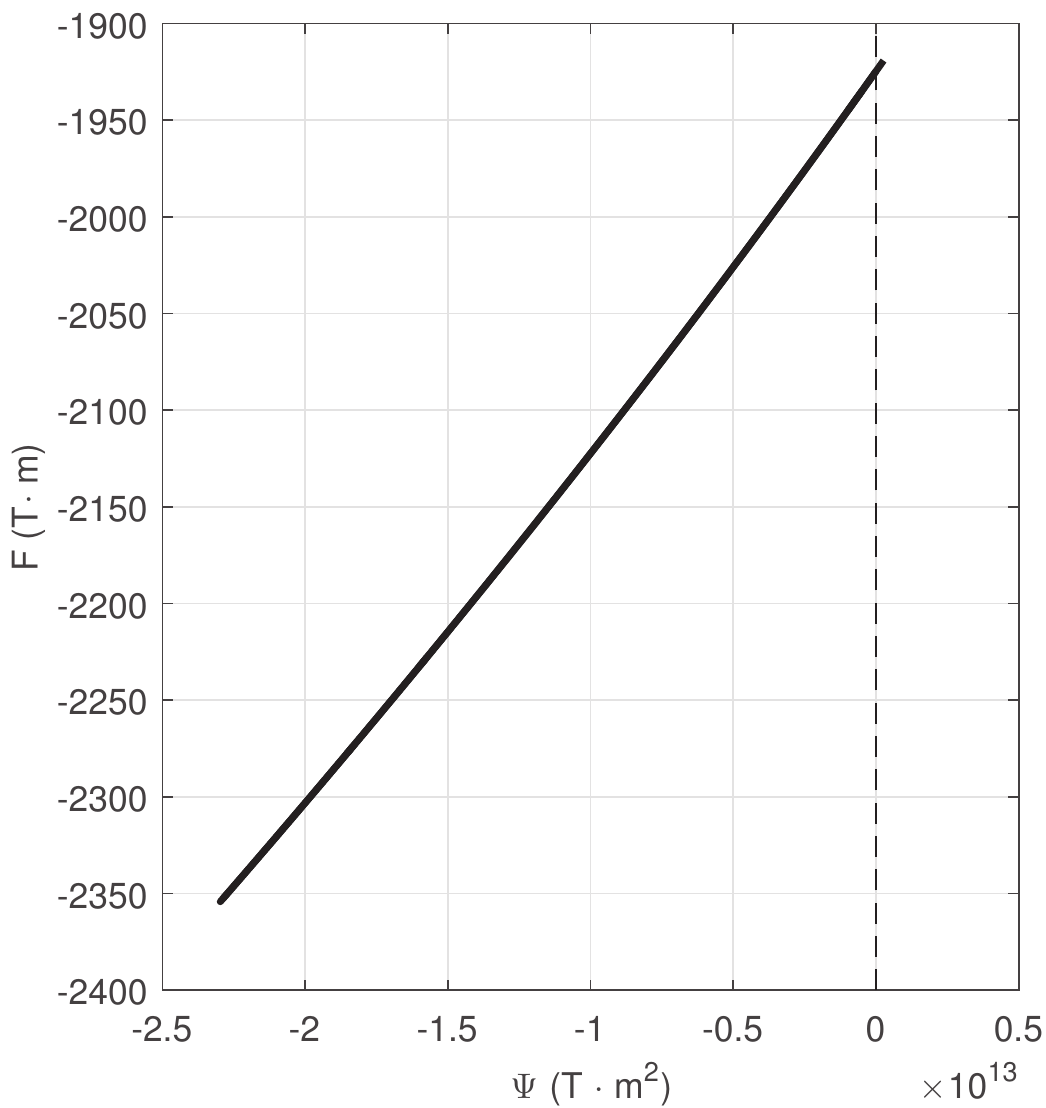}
 }
         \vspace{-0.50\textwidth}   % Shift close to the panel top
     \centerline{ \bf     % Includes the labels (here needs the color package)
      \hspace{0.05 \textwidth} {(a)}
      \hspace{0.45\textwidth}  {(b)}
         \hfill}
     \vspace{0.50\textwidth}    % Shift back to the panel bottom
\caption{The corresponding measured $F$ versus $\Psi$ data points
along the spacecraft path, and the 2nd-order polynomial fitting
 $F(\Psi)$ (black curve) for (a) the numerical GS solution.
 A fitting residue $R_f$ and a flux rope boundary $\Psi=\Psi_b$ are also marked \citep{2004JGRAHu}.
 The corresponding exact $F(\Psi)$ function is shown in (b).}\label{fig:pta003}
 \end{figure}

Figure~\ref{fig:chi2} shows the determination of the location
of the $Z$ axis as the intersection of $Z$ with the $(r_{sc},t)$
plane through the $\chi^2$ minimization procedures after the first
step when an optimal $Z$ axis orientation is chosen from the
residue map (see Paper I). The advantage of such an approach with
measurement uncertainties is to enable proper $\chi^2$ statistics in order
to provide an objective assessment of the goodness-of-fit
\citep{2002nrca.book.....P}. A reduced $\chi^2$ value $\approx 1$
represents the range of uncertainty for the optimal parameter to
be determined, \textit{i.e.} the location of $Z$ on the $(r_{sc},
t)$ plane within certain domain as illustrated in
Figure~\ref{fig:chi2}. The independent assessment of the
goodness-of-fit is obtained by calculating the quantity $Q$
 \citep[][see also, Paper
I]{2002nrca.book.....P}, as shown in Figure~\ref{fig:chi2}b,
associated with each search grid point. Such a quantity indicates the probability of a value drawn from a $\chi^2$ distribution greater than the specific $\chi^2$ value. Usually,  a large $Q$ value close
to 1, together with reduced $\chi^2\lesssim 1$, indicates a 
``modestly" good fit \citep{2002nrca.book.....P}. We use the
combined criteria of $\chi^2\le 1$ and $Q\ge 0.9$ as the range of
uncertainty for selecting the optimal parameters.

In this case, the regions of uncertainty  given by both contours
of $Q=0.9$ and $\chi^2=1$ nearly overlap on the $(r_{sc},t)$
plane. We choose the location marked by the asterisk near the
center of the closed contour $\chi^2=1$ as the optimal choice of
the location of the $Z$ axis to carry out the final
reconstruction procedures. The resulting major radius is
$r_0=0.92$ AU, which deviates from the true value, $r_0=1.02$ AU, by about 10\%.
\begin{figure}
 \centerline{\includegraphics[width=0.5\textwidth,clip=]{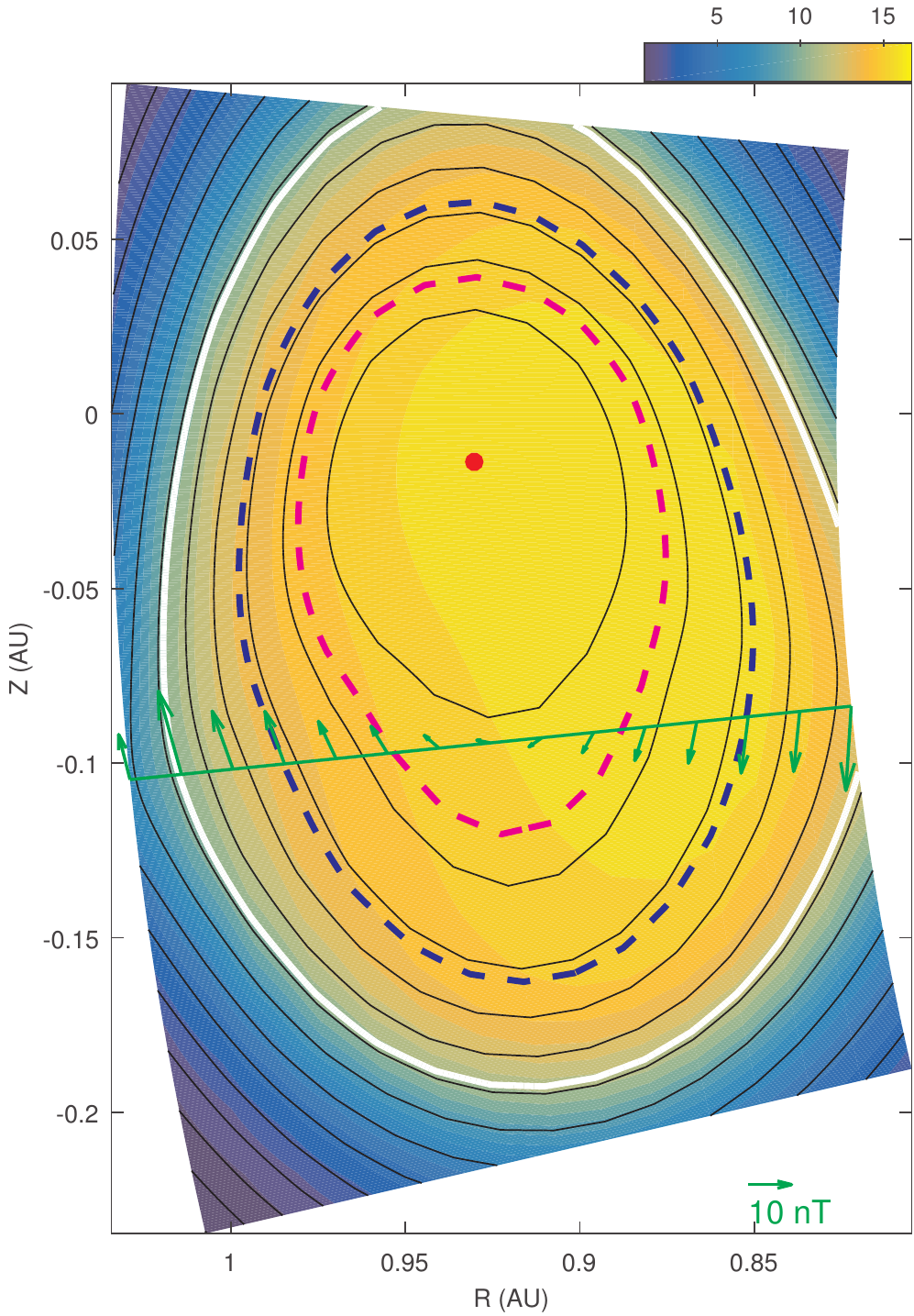}
 \includegraphics[width=0.5\textwidth,clip=]{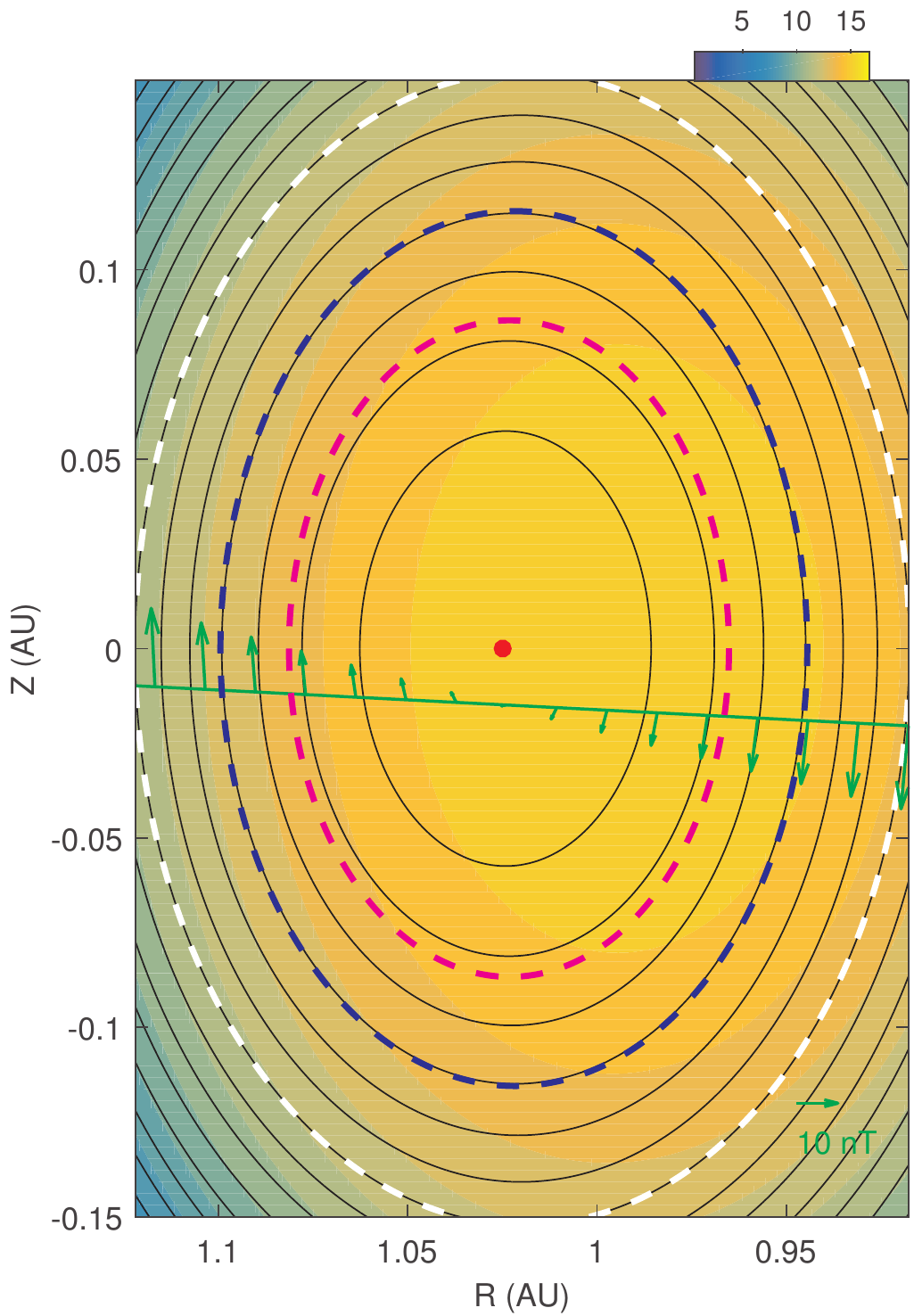}
 }
%        \vspace{-0.42\textwidth}   % Shift close to the panel top
%     \centerline{ \bf     % Includes the labels (here needs the color package)
%      \hspace{0.06 \textwidth} \color{white}{(a)}
%      \hspace{0.44\textwidth}  \color{white}{(b)}
%         \hfill}
%     \vspace{0.42\textwidth}    % Shift back to the panel bottom
 \caption{The resulting numerical GS solution (left panel) and the exact solution (right panel) for the benchmark case.
 In each panel, the black contour lines are equi-value contours of $\Psi$, and the filled color contour represents the axial field
 component
 $B_\phi$ with the same range of scales indicated by the colorbars (positive being out of the plane). The red dot and color dashed contours  are the selected field lines of
 approximately the same $\Psi'$ values, among which the red dot denotes the location where the transverse field vanishes.  Additionally, the green arrows represent the measured transverse magnetic field components along
 the projected spacecraft path (the green line). The white contours represent $\Psi=0$ (dashed) and $\Psi=\Psi_b$, respectively. }\label{fig:map003}
 \end{figure}

Now we complete the last steps of the GS reconstruction by
applying the GS solver described in Paper I to obtain a numerical
solution of the GS equation, using the geometrical parameters
determined above. These are additional steps not carried out in
Paper I. All together, a complete real event study will start with
the two-step recipe described in Paper I, followed by these last
steps. Figure~\ref{fig:pta003} shows the resulting $F$
\textit{versus} $\Psi$ data plot along the spacecraft path and the
corresponding fitting curve with a fitting residue $R_f=0.11$
defined as before \citep{2004JGRAHu}. Figure~\ref{fig:map003}
shows the cross sectional map of the numerical GS solution
compared with the exact one. Since these two solutions are no
longer lying on the same plane, due to different $Z$ axis
orientations and different $r_0$ values, a quantitative evaluation
of error through point-by-point subtraction is no longer valid.
The two solutions clearly differ in overall shape, size and
particularly the impact parameter, $d_0$, the shortest distance
between the center of the flux rope (the red dot) and the projected
spacecraft path (the green line). The overall axial field profiles are
similar, but only within the boundary $\Psi=\Psi_b$ of the
numerical solution and that $\Psi=0$ of the exact solution. Both
are of left-handed chirality. The axial field at the center is 16
nT and 15 nT, respectively, for the numerical and exact solution.

\begin{figure}
 \centerline{\includegraphics[width=0.45\textwidth,clip=]{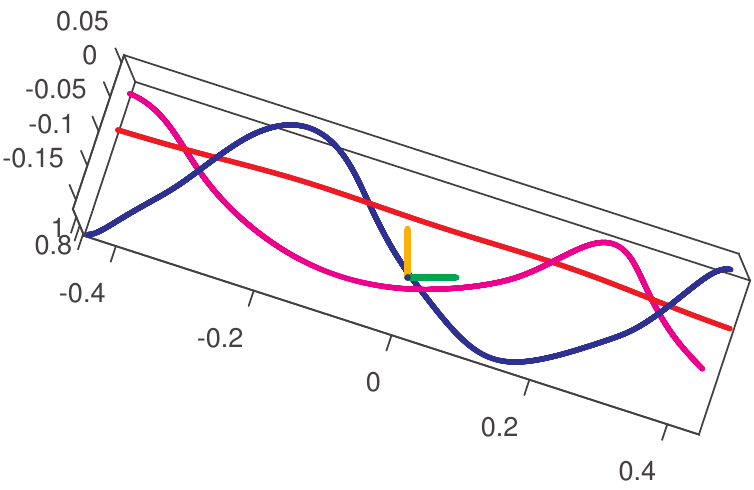}
 \includegraphics[width=0.55\textwidth,clip=]{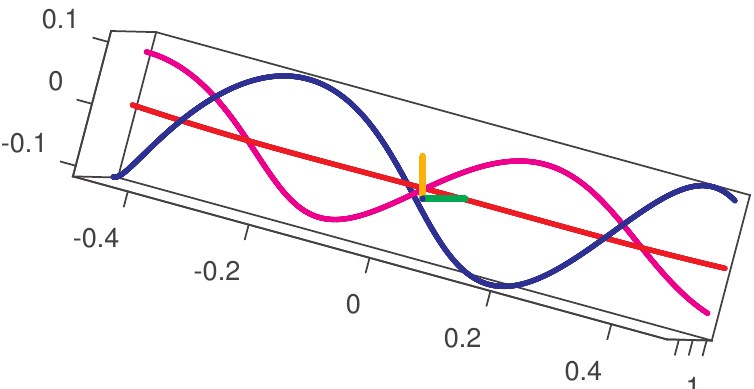}
}
%         \vspace{-0.50\textwidth}   % Shift close to the panel top
%     \centerline{ \bf     % Includes the labels (here needs the color package)
%      \hspace{0.05 \textwidth} {(a)}
%      \hspace{0.45\textwidth}  {(b)}
%         \hfill}
%     \vspace{0.50\textwidth}    % Shift back to the panel bottom
\caption{The 3D view against the radial direction, $r_{sc}$, of
selected field lines of the same $\Psi'$ values for both the
numerical (left panel) and  exact (right panel) GS solutions. The
short thick lines are the $r_{sc}$ (blue), $t$ (green), and $n$
(gold) unit vectors, respectively, located at the point of
spacecraft entry into the flux rope interval.}\label{fig:3dR}
 \end{figure}

\begin{figure}
 \centerline{\includegraphics[width=0.45\textwidth,clip=]{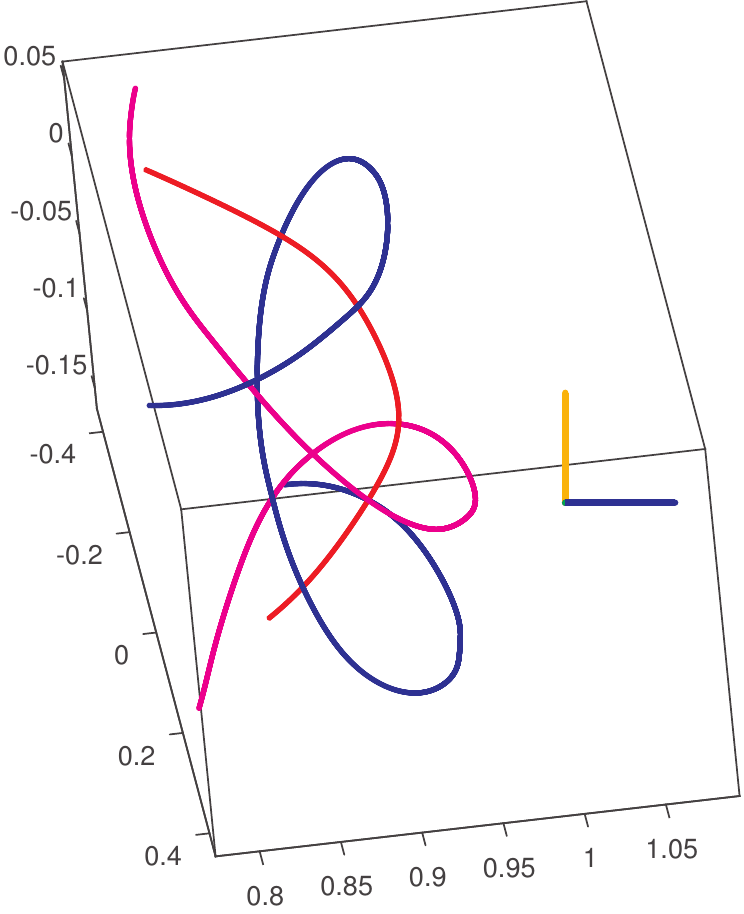}
 \includegraphics[width=0.55\textwidth,clip=]{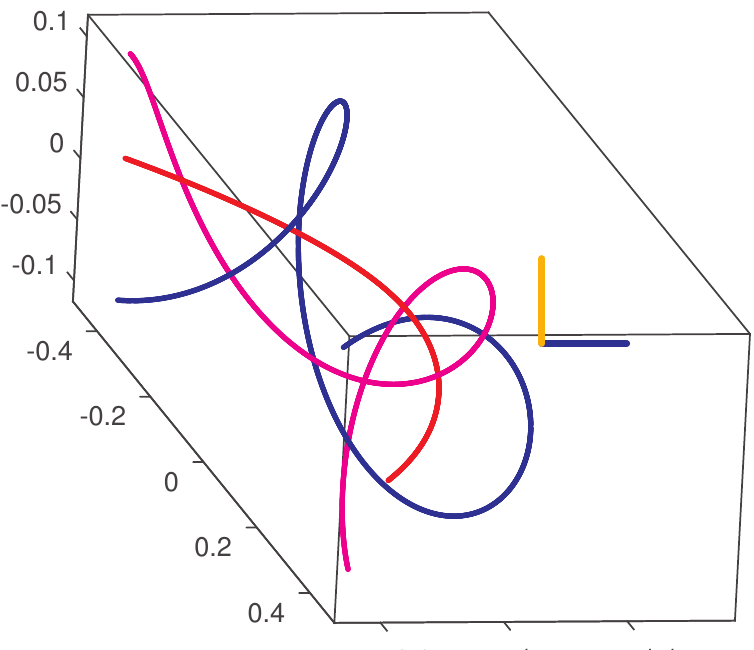}
 }
%         \vspace{-0.50\textwidth}   % Shift close to the panel top
%     \centerline{ \bf     % Includes the labels (here needs the color package)
%      \hspace{0.05 \textwidth} {(a)}
%      \hspace{0.45\textwidth}  {(b)}
%         \hfill}
%     \vspace{0.50\textwidth}    % Shift back to the panel bottom
\caption{The solutions in  3D view along the  direction $t$ with
$n$ upward. The format is the same as in Figure~\ref{fig:3dR}.
}\label{fig:3dT}
 \end{figure}

\begin{figure}
 \centerline{\includegraphics[width=0.48\textwidth,clip=]{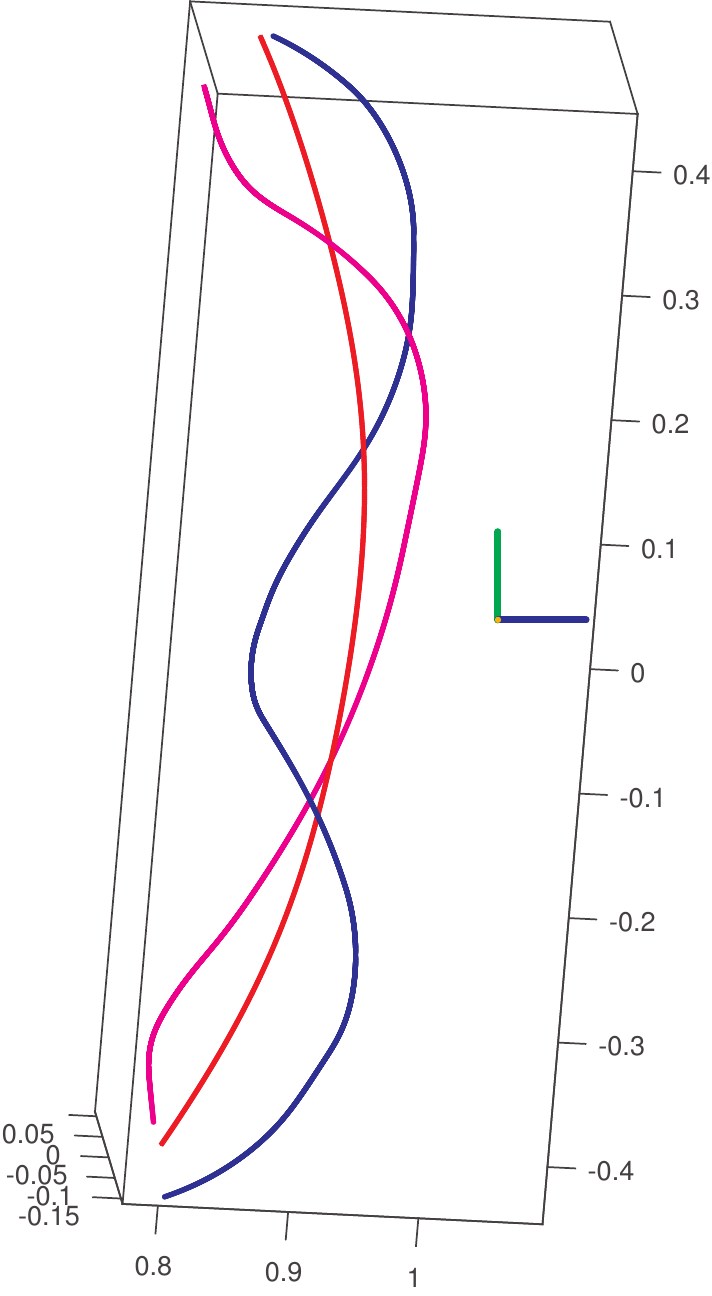}
 \includegraphics[width=0.52\textwidth,clip=]{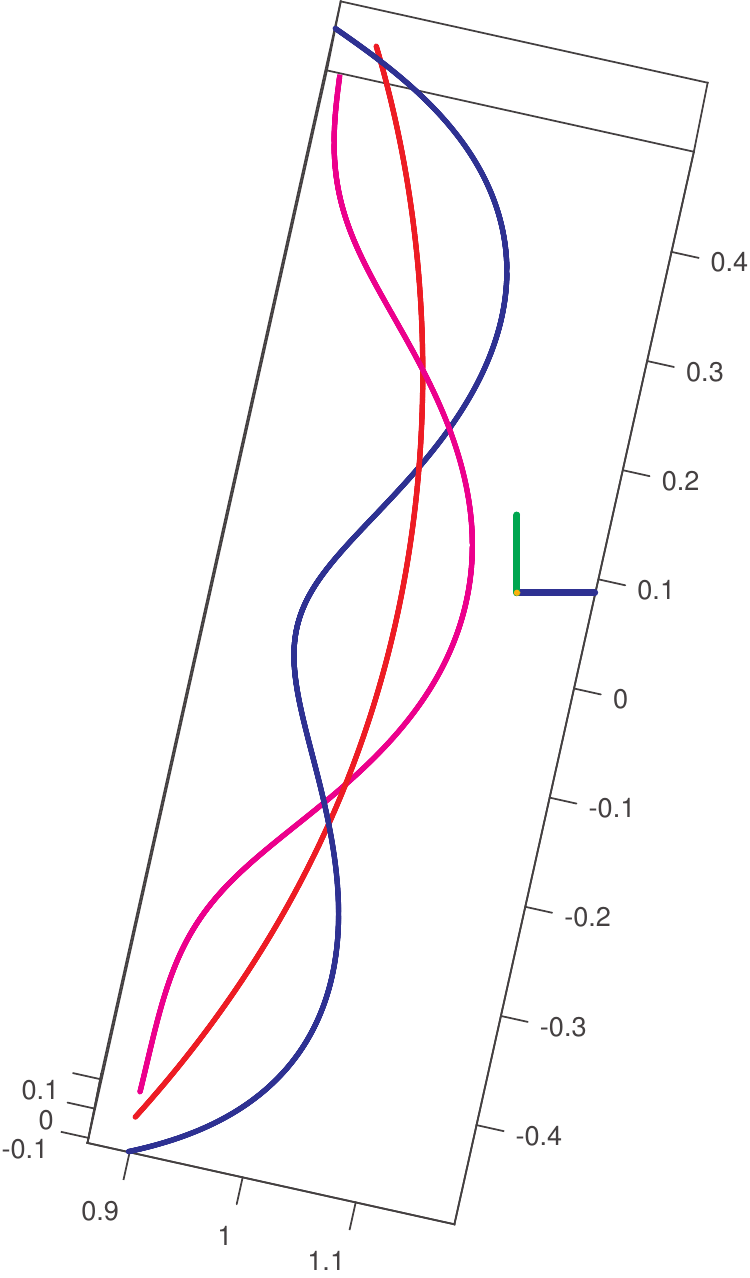}
 }
%         \vspace{-0.50\textwidth}   % Shift close to the panel top
%     \centerline{ \bf     % Includes the labels (here needs the color package)
%      \hspace{0.05 \textwidth} {(a)}
%      \hspace{0.45\textwidth}  {(b)}
%         \hfill}
%     \vspace{0.50\textwidth}    % Shift back to the panel bottom
\caption{The solutions in  3D view against the  direction $n$. The
format is the same as in Figure~\ref{fig:3dR}. }\label{fig:3dN}
 \end{figure}

To further compare the magnetic field configuration between these
two solutions, we present the 3D field-line plots of three
selected field lines of similar $\Psi'=\Psi-\Psi_0$ values, where
the flux function value at the center is denoted $\Psi_0$, for
both solutions. Their projections onto the cross sectional planes
are shown in Figure~\ref{fig:map003} as nested closed loops by
colored dashed lines including the red dots. They can be
considered as the same set of field lines from the two solutions,
thus can be  inter-compared. Such comparisons are shown in
Figures~\ref{fig:3dR}, \ref{fig:3dT}, and \ref{fig:3dN}, for three
different view angles, respectively. The discrepancies in their
overall shapes are seen, albeit some field lines, for example, the
central red and the outer blue ones appear to have better
agreement between the numerical  and the exact GS solutions.

 \begin{table}
 \caption{Comparison of the outputs between the numerical ($R_f=0.11$) and exact $(R_f=0.0)$ GS  solutions. }\label{tbl:bench}
 \begin{tabular}{ccccc}
 \hline
 %\multicolumn{2}{c}{<>}
 $R_f$ & $B_{\phi,0}$ (nT) & $\Phi_p$ ($10^{12}$Wb/radian) &
$\Phi_t$ ($10^{12}$Wb) & $d_0$ (AU) \\
\hline
0.0 & 15 & 23 & 15 & 0.015 \\
0.11 & 16 & 23 & 14 & 0.080 \\\hline
 \end{tabular}
 \end{table}
In addition, we examine the quantitative output of some bulk
properties derived  from the GS reconstruction result.
Table~\ref{tbl:bench} summarizes the comparison of various output
parameters between the numerical GS toroidal reconstruction and
the exact solution. They include the central axial field
$B_{\phi,0}$, the poloidal and axial magnetic  flux, $\Phi_{p,t}$, and the
impact parameter $d_0$. Despite the large deviation in the impact
parameter as we pointed out earlier, the other parameters,
including the magnetic flux content, agree between the numerical
GS reconstruction result and the exact solution.

\section{Application to \textit{In Situ} MC Events}\label{sec:events}

%\subsection{STEREO B Event: 6 June 2008}
We present here a complete real MC event study by the toroidal GS
reconstruction. The event was observed by the STEREO B (STB)
spacecraft near 1~AU on 6-7 June 2008. This event was also
examined by both the original  straight-cylinder GS
reconstruction, using the \textit{in situ} STB data in
\citet{2009ApJM}, and by the white-light imaging reconstruction
of the flux rope evolution through the inner heliosphere in
\citet{2010ApJ...715.1524W}. Both studies focused on exploiting
the remote-sensing observations and morphological  modeling, which
provides continuous coverage of CME propagation and morphology of
the modeled flux rope structure en route from the Sun to 1~AU
\citep[see also,][]{2012SoPhW}. The current study contributes to
the goal of linking remote-sensing observations with \textit{in
situ} measurements and modeling, especially in addressing the
magnetic field configuration of the flux rope. For this particular
event, an inter-comparison between these studies is  available.

% add a time-series plot
\begin{figure}
 \centerline{\includegraphics[width=.8\textwidth,clip=]{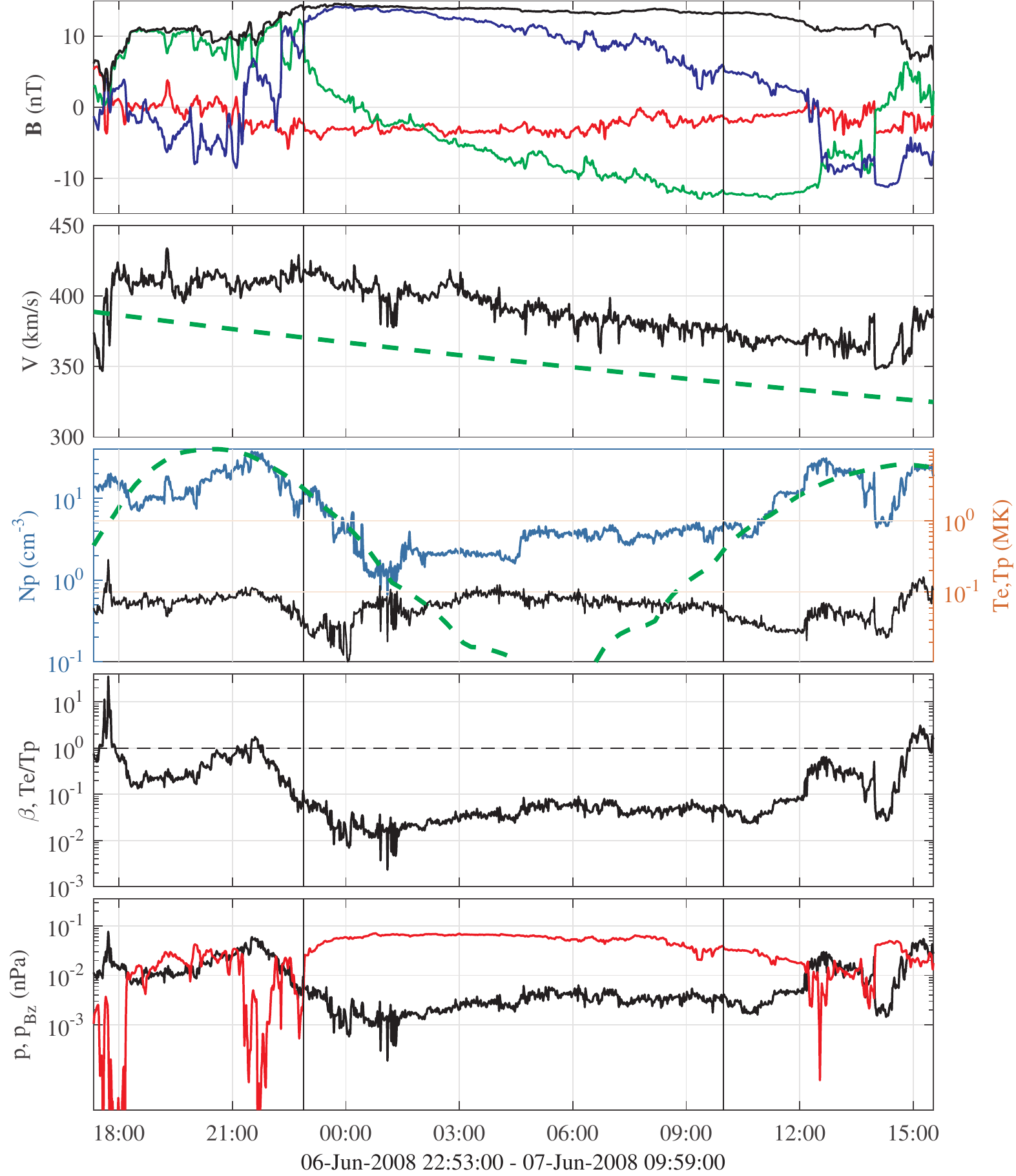}}
 \caption{The time series from STB \textit{in situ} measurements. From top to bottom panels: the magnetic field components in the spacecraft
 centered $(r_{sc},t,n)$ coordinates and the magnitude in red, green,  blue, and black, respectively, the solar wind speed, the proton
 density (left axis) and temperature (right axis; the electron temperature $T_e$ not available), the proton $\beta$ (the dashed horizontal line marks the value 1.0), and the plasma
 and axial magnetic pressure (red). The dashed green curves in the second and the third panels are the corresponding model predictions from \citet{2010ApJ...715.1524W}.
 The vertical lines mark the GS reconstruction interval selected for this study, given beneath the plot.  }\label{fig:data}
 \end{figure}
 Figure~\ref{fig:data} shows the time-series plot of both magnetic
field and plasma parameters measured at STB. The GS reconstruction
interval for the toroidal geometry is selected as marked by the
vertical lines. This interval corresponds well to a region of
depressed proton $\beta$ (electron temperature is not available)
and elevated magnetic field magnitude. In addition, the speed and
density profiles obtained from Wood et al.'s imaging
reconstruction analysis are also superposed, indicating a good match in
the density peaks enclosing the flux rope interval. These peaks
correspond to the outer surface  of the flux rope volume from
white-light imaging reconstruction, as visualized in
\citet{2010ApJ...715.1524W} (see also Figure~\ref{fig:3DSTB}).

% add the residue map here
\begin{figure}
 \centerline{\includegraphics[width=.6\textwidth,clip=]{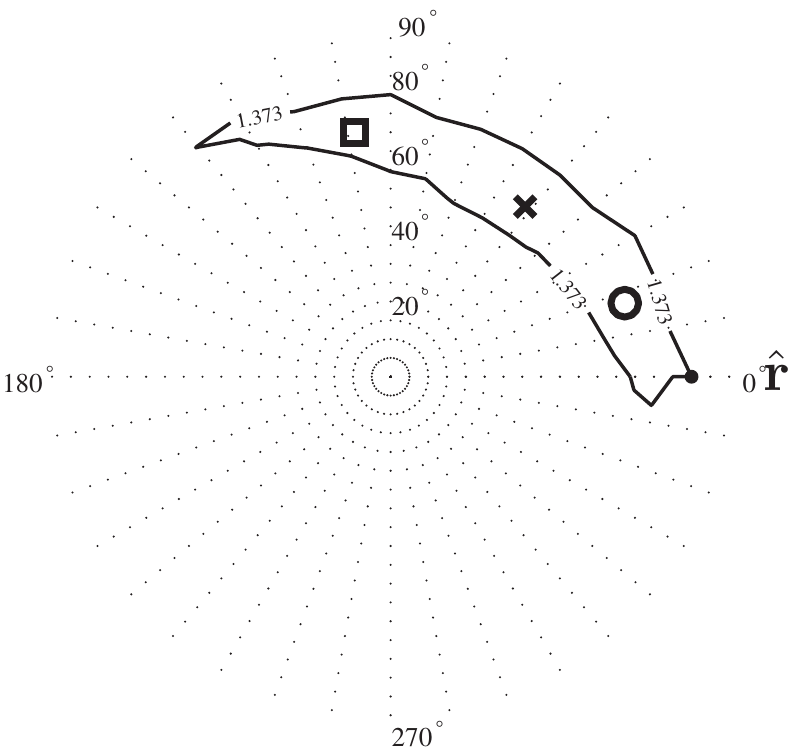}}
 \caption{The residue map of the STB event: the contours of $Res$. The background dots are tips of unit vectors  over the hemisphere of unit radius, enumerating all possible directions in space. The directions at  0 and 90 degrees correspond to unit vectors along $r_{sc}$ and $t$, respectively. The  cross sign marks the optimal rotation axis $Z$ direction (see Paper I). The other two symbols mark alternatives
 inside the  contour of value $1+\min({Res})$. }\label{fig:resmap}
 \end{figure}
The toroidal GS reconstruction is then performed, following the
procedures laid out in Paper~I and in the previous section. A
standard set of plots and outputs is to be presented in a
sequential order. Figure~\ref{fig:resmap} displays the usual
residue map, contours of residue, $Res$ (see Paper I for
definition), over a search grid of trial $Z$ axis. This results  from the first step of the two-step recipe for
determining the orientation of $Z$ through a trial-and-error
process by enumerating all possible directions of $Z$. The thick
black contour is of value, $Res=\min(Res)+1$, representing the
range of possible $Z$ axis orientations. Based on the recipe and
the benchmark study in Paper I,  the optimal $Z$ axis orientation
is usually chosen  as the point near the center of the contour, as
marked by the cross sign, for such a singly connected contour. It
is $Z=(0.5239,0.6626,0.5353)$ in the $(r_{sc},t,n)$ coordinates
and is used for the subsequent reconstruction. Optionally we also
denote two other possible choices enclosed by the contour, which
introduce significant uncertainty in the reconstruction result,
but can be ruled out by other observations and associated analysis
to be discussed later (see also Table~\ref{tbl:2}).

\begin{figure}
 \centerline{\includegraphics[width=1.\textwidth,clip=]{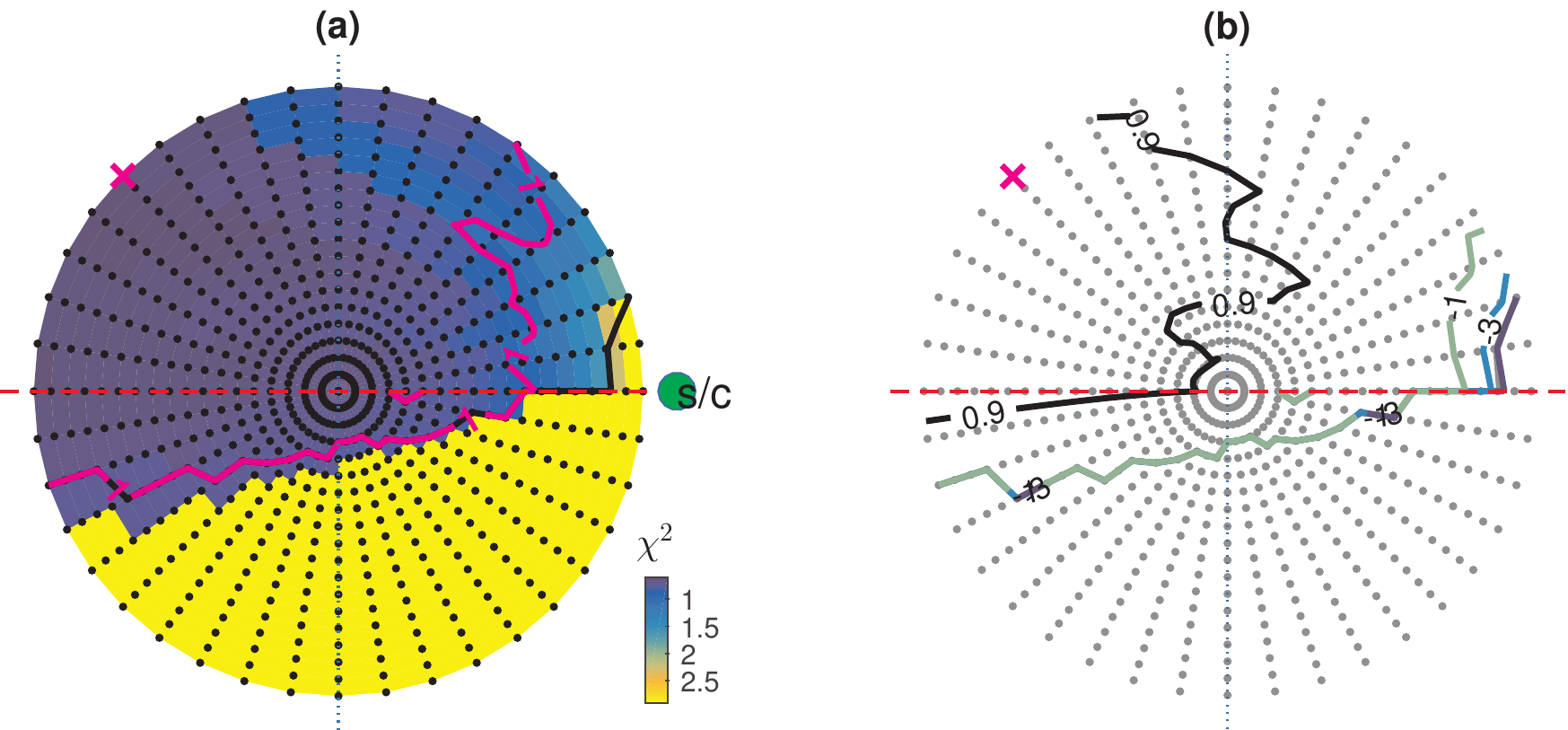}}
 \caption{(a) The distribution of reduced $\chi^2$ value, as indicated by the colorbar  on the $(r_{sc},t)$ plane.
  (b) The corresponding contour plot of $\log_{10}Q$ as labeled except that the thick black contour is of level $Q=0.9$, and the outermost one is the same as the black one in (a). The format is the same as in Figure~\ref{fig:chi2}.
    The  cross sign
 marks the  location  of $\chi^2_{\min}$.}\label{fig:chi2STB}
 \end{figure}
\begin{figure}
 \centerline{\includegraphics[width=.5\textwidth,clip=]{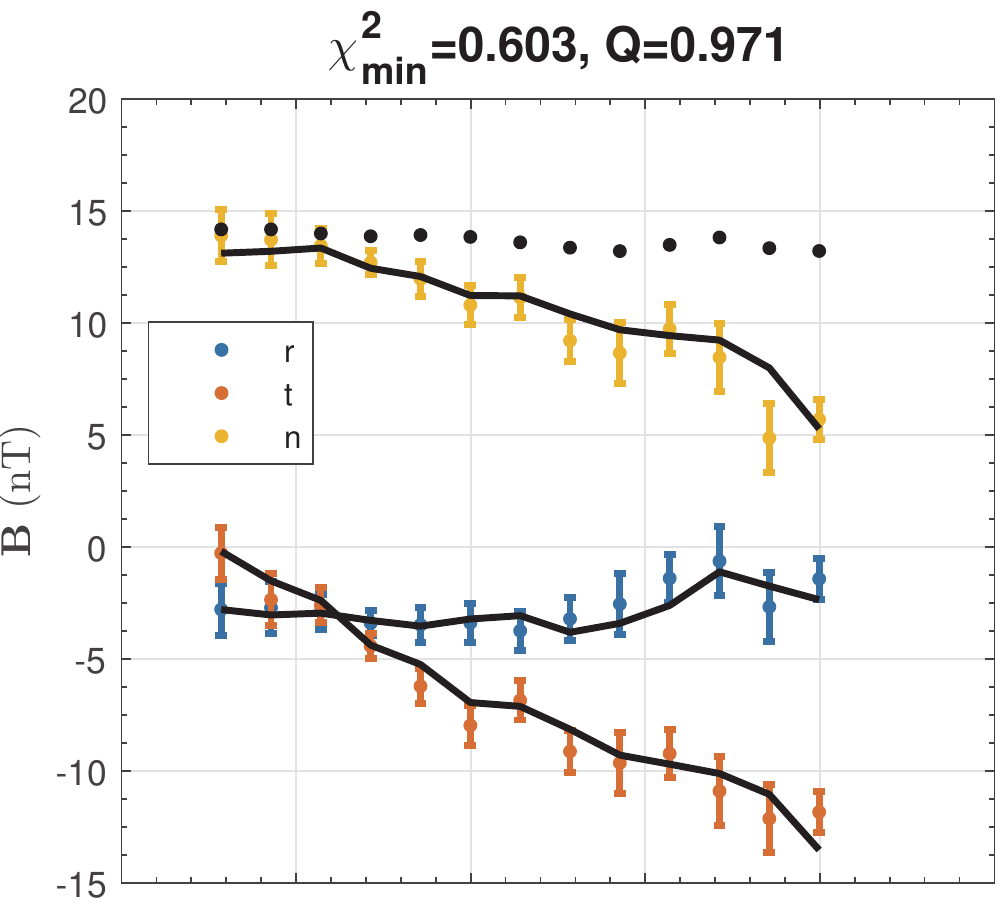}}
 \caption{The magnetic field components (colored dots; see legend) along the spacecraft path with uncertainties (errorbars)
  and the corresponding GS model output (black lines) for the STB MC event. The associated minimum reduced $\chi^2$  and
  $Q$ values are given on top.}\label{fig:Bxyz}
 \end{figure}
\begin{figure}
 \centerline{\includegraphics[width=.45\textwidth,clip=]{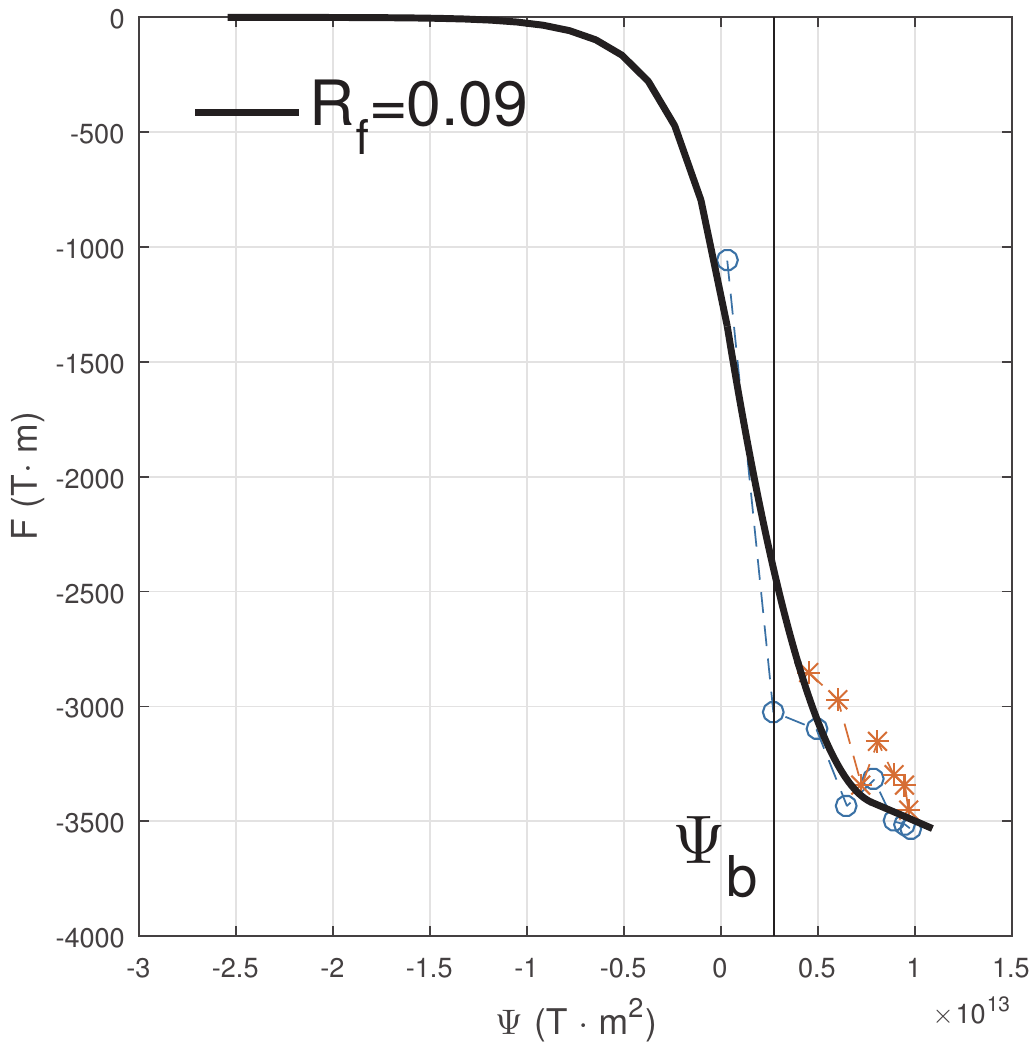}
 \includegraphics[width=.55\textwidth,clip=]{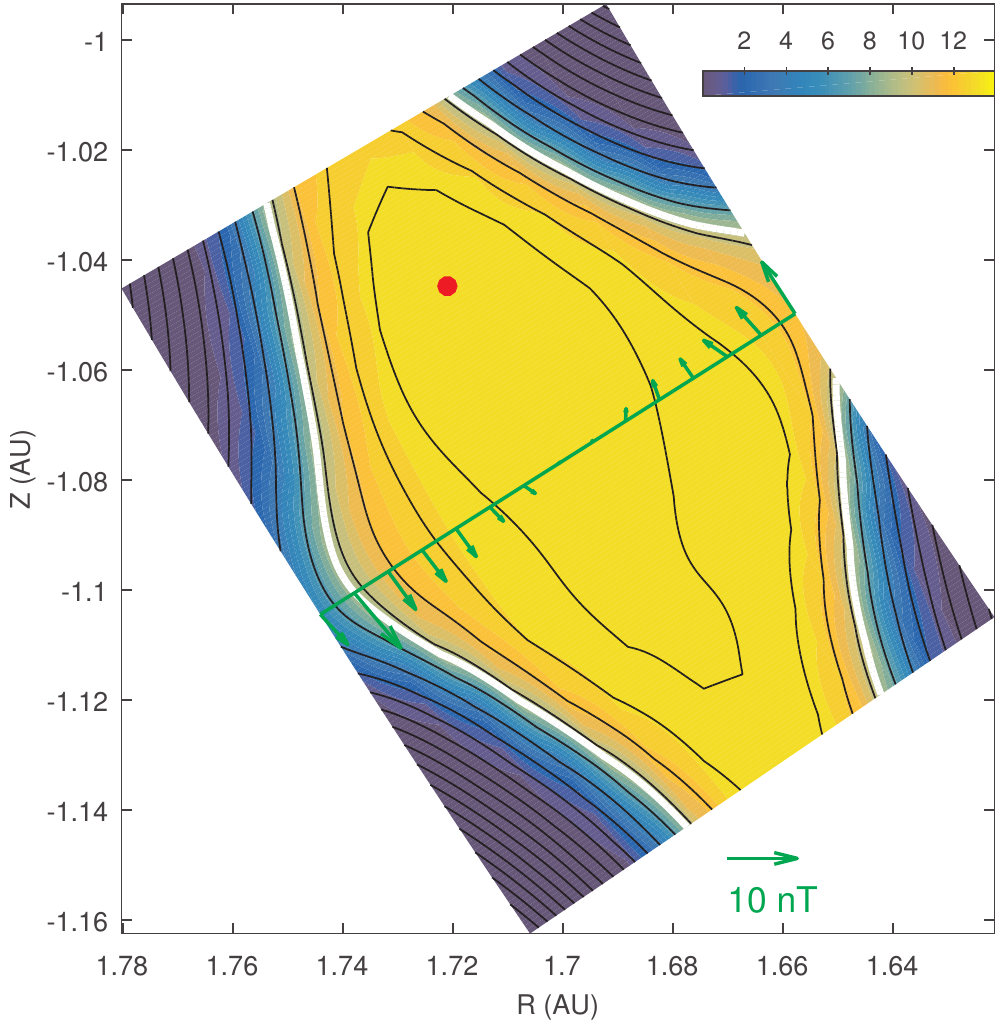}}
         \vspace{-0.10\textwidth}   % Shift upward
     \centerline{ \bf     % Includes the labels (here needs the color package)
      \hspace{0.05 \textwidth} {(a)}
      \hspace{0.45\textwidth}  {(b)}
         \hfill}
     \vspace{0.10\textwidth}    % Shift back to the panel bottom
      \caption{The toroidal GS reconstruction result for the STB MC event. (a) The composite function $F$ \textit{versus}
      $\Psi$ and the fitting curve. The format is the same as in Figure~\ref{fig:pta003}. (b) The resulting cross section of the
      toroidal flux rope on the $(R,Z)$ plane. The format is the same as in Figure~\ref{fig:map003}. }\label{fig:Ptmap}
 \end{figure}
\begin{figure}
 \centerline{\includegraphics[width=.4\textwidth,clip=]{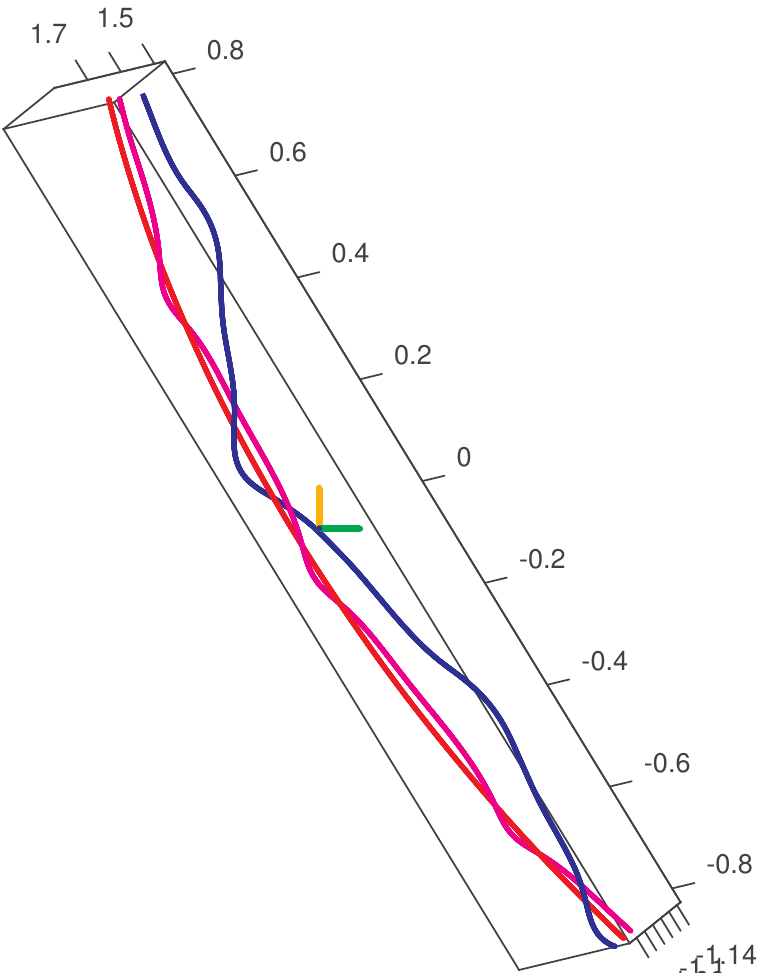}
 \hspace{.1\textwidth}
 \includegraphics[width=.5\textwidth,clip=]{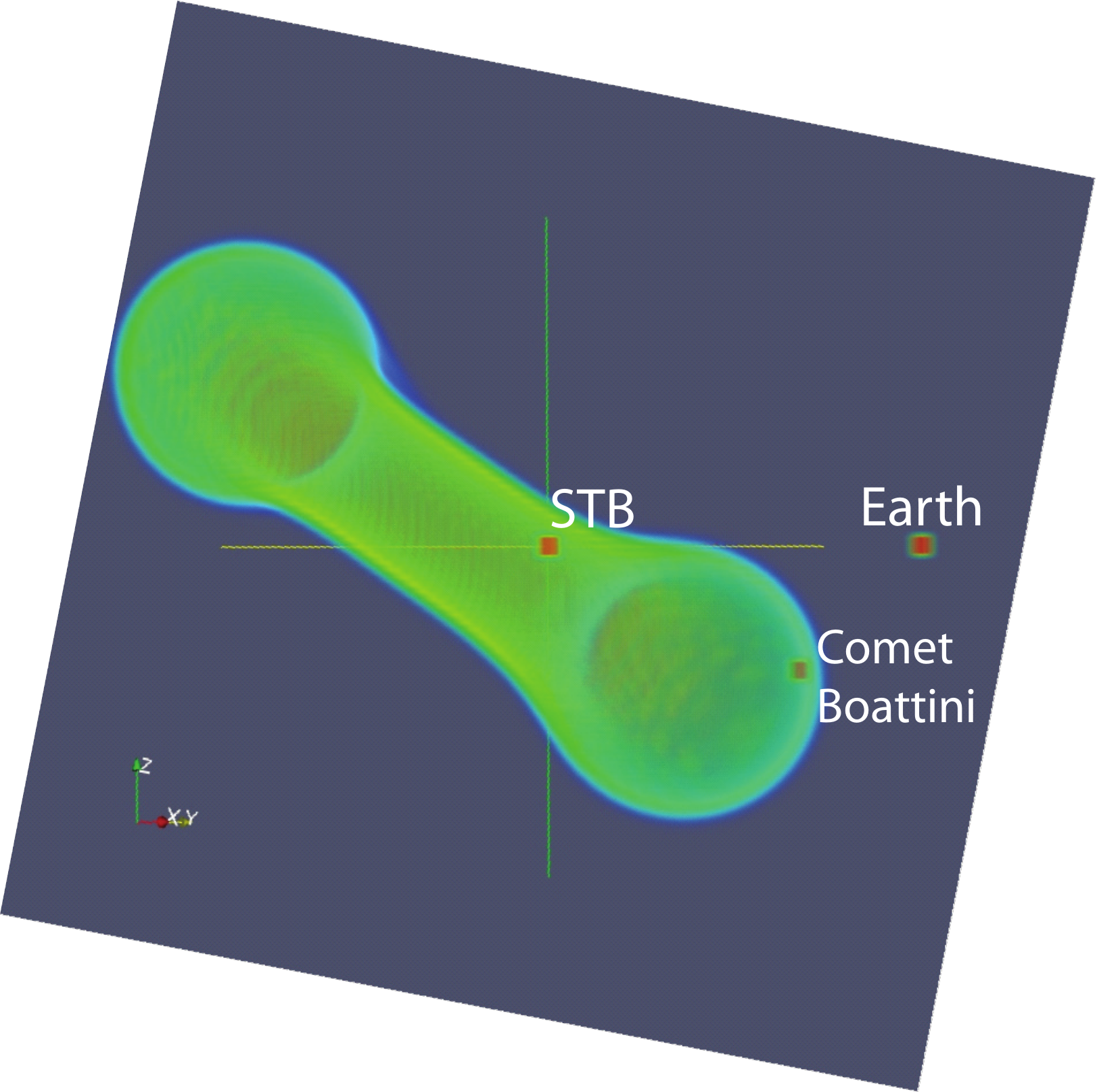}}
 \caption{Left panel: the 3D view based on the reconstruction result of Figure~\ref{fig:Ptmap}b. The view point
 is toward the Sun. The format is the same as in Figure~\ref{fig:3dR}. Right panel:
 the flux rope configuration obtained from \citet{2010ApJ...715.1524W} as seen from the same view angle and at the same location and time as the left panel. The STB location (the center dot) corresponds to the intersection of the short green and gold lines in the left panel. }\label{fig:3DSTB}
 \end{figure}

 Figure~\ref{fig:chi2STB}, in the same format as in
 Figure~\ref{fig:chi2}, demonstrates the determination of the
 other important geometrical parameter, the major radius $r_0$, or
 the intersection of the $Z$ axis with the $(r_{sc},t)$ plane.
 This result is obtained from the second step of the two-step
 recipe of Paper I. Significant  amount  of uncertainty exists for the combined
 criteria of $\chi^2\le 1$ and $Q\ge 0.9$ in determining an
 optimal solution of $r_0$. The range of uncertainty in $r_0$ is
 reflected by the area enclosed by the thick black contour in
 Figure~\ref{fig:chi2STB}b, which may extend to an even larger
 region beyond the search grid shown. For such a region enclosed by
 the $Q=0.9$ contour, the corresponding ranges of certain parameters
 including the major radius are given in Table~\ref{tbl:2}. We
 choose the optimal $Z$ axis location at the point of minimum
 reduced $\chi^2$ value, $\chi^2=\chi^2_{\min}$, as marked by the
 magenta cross sign.  This choice is considered representative,
 given that both the $\chi^2=1$ and $Q=0.9$ contour lines are open
 in this case. The resulting major radius is $r_0=1.70 $ AU. The
 corresponding reduced $\chi^2$ and $Q$ values are 0.603 and
 0.971, respectively, as indicated in Figure~\ref{fig:Bxyz}. There
 the measured  magnetic field components and the output from GS
 model fitting of $F=RB_\phi$ \textit{versus} $\Psi$ along the
 spacecraft path are shown, together with measurement
 uncertainties associated with each data point. Such
 uncertainties, $\sigma_i$,
 represented by errorbars associated with each point, are
 estimated from higher-resolution measurements by a calculation of
 the root-mean-squres of each magnetic field component $B_\nu$ within certain segment of $M$ data points with higher resolution, \textit{e.g. via}  $\sigma_\nu^2=\sum_{i=1}^M\left\langle(B_{\nu i}-\langle B_\nu\rangle)^2\right\rangle$ \footnote{See the Ace Science
 Center: \url{http://www.srl.caltech.edu/ACE/ASC/level2/mag_l2desc.html}, the calculation of the parameter, \tt{dBrms}.}. The measured quantities $F$ and $\Psi$ along the
 spacecraft path, and the corresponding fitting curve $F(\Psi)$
 are given in Figure~\ref{fig:Ptmap}a. The final reconstructed
 cross section in an annular computation domain is shown in
 Figure~\ref{fig:Ptmap}b, using the toroidal GS solver described
 and tested in Paper~I.

%\begin{figure}
% \centerline{\includegraphics[width=.5\textwidth,clip=]{STB_allSTB158r08_2.pdf}}
% \caption{}\label{fig:Ptall}
% \end{figure}

The cross sectional map shows an elongated shape with fairly
constant axial field, $B_\phi$, and a flux rope configuration of
right-handed chirality. The center of the flux rope, defined as
the location where the transverse field vanishes, locates at a
distance $d_0=0.038$ AU away from the spacecraft path. At this location
(the red point), the field line becomes ``straight", \textit{i.e.}
it forms an exact circle around the $Z$ axis, concentric with the
circle of major radius $r_0$. To further illustrate the
configuration of such a toroidal flux rope, a 3D view with
selected field lines is shown in Figure~\ref{fig:3DSTB} (left
panel) with  a view angle radially toward the Sun from the STB point
of view. The central field line in red is rooted on the red dot in
Figure~\ref{fig:Ptmap}b, while the other two field lines are
winding around this central axis of a finite curvature.
Figure~\ref{fig:3DSTB} (right panel), reproduced from
\citet{2010ApJ...715.1524W}, shows the same view of the
reconstructed flux rope from white-light images, reaching STB's
location at 1 AU. The colored surface represents the boundary of
the flux rope, corresponding to the peak density values in
Figure~\ref{fig:data}. The reconstruction was validated by its success in reproducing the observed time or arrival at STB, and also its observed time of encounter with a comet that happened to be nearby (Comet Boattini).  The inclination angle of the apparent
central axis of the flux rope as seen away from the ecliptic plane
is about 35 degrees (equivalent to a clock angle $\phi_t=145$
degrees, to be defined below).

\begin{figure}
 \centerline{\includegraphics[width=1.\textwidth,clip=]{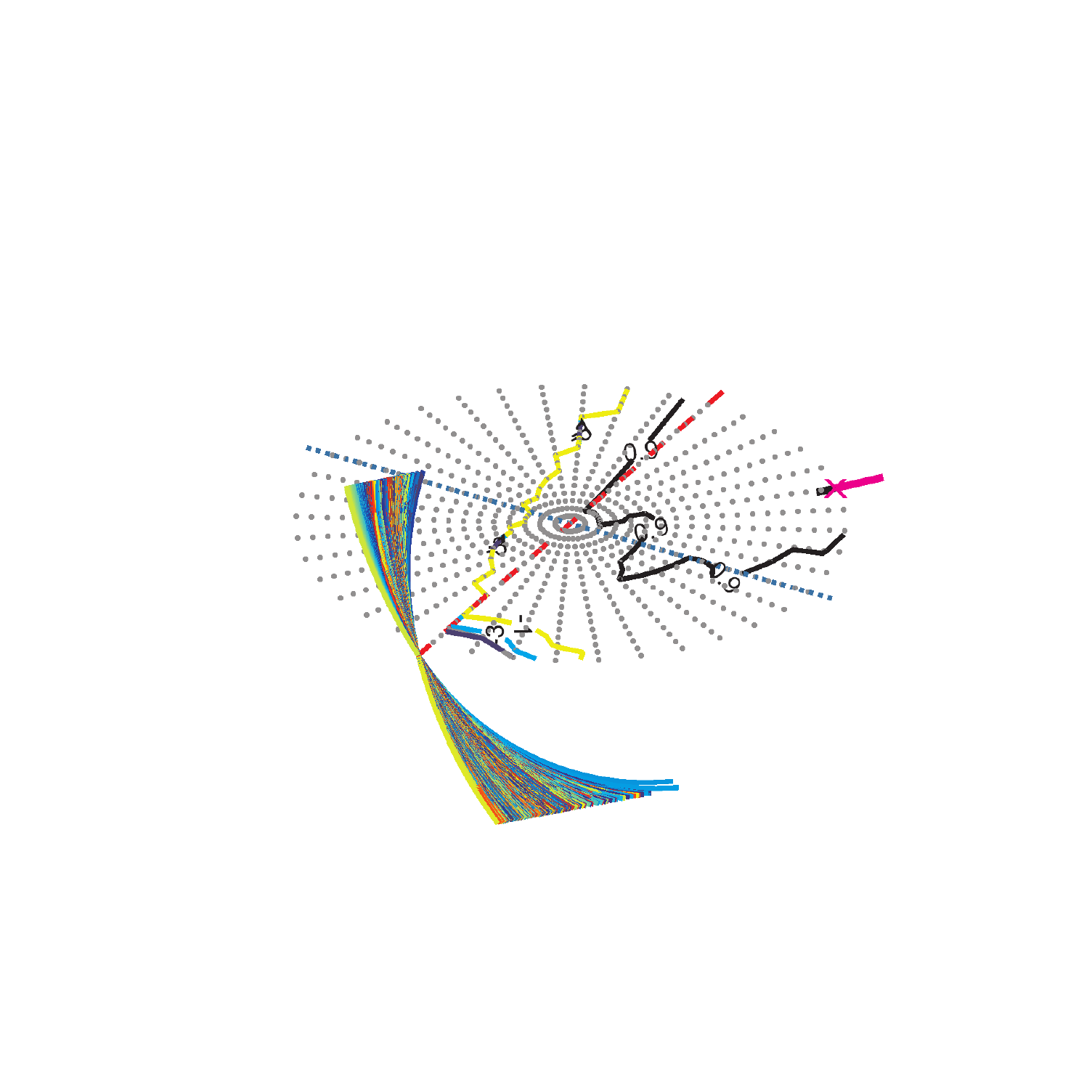}}
 \caption{Figure~\ref{fig:chi2STB}b rendered in a different view angle. The red dashed line and the blue dotted line are
 the $r_{sc}$ and $t$ coordinates, respectively. Additionally, the thick pink line and the cross sign mark the direction and location of
 optimal $Z$ axis, the orientation of which is also marked  on Figure~\ref{fig:resmap}. The bunch of arcs all crossing the spacecraft location at $r_{sc}=r_{sc0}$
 illustrate the sections of circles centered around the $Z$ axis whose location changes over the dots enclosed by the
 thick black contour ($Q=0.9$) while maintaining its direction. Each colored arc is the result from the corresponding
  location in this case.}\label{fig:3Darcs}
 \end{figure}

%
 %
%  -0.134191535984240   0.901417056937613   0.411630806830736
Figure~\ref{fig:3Darcs} attempts to illustrate the uncertainty in
the geometry of the reconstructed  toroidal flux rope configuration at the
STB location due to the uncertainty in the location of the
central rotation axis $Z$ while keeping its orientation fixed. The
set of arcs represents the curvature of the toroidal flux rope at
the point of spacecraft intersection. We define two angles
associated with the directional vector tangent to each arc at the
point of spacecraft location, namely, the cone angle $\phi_r$, the
angle between the directional vector and the radial direction
$r_{sc}$, in the range [0, 180] degrees, and the clock angle
$\phi_t$, the angle of the projection of the tangent vector  on
the $(t,n)$ plane, measured from the positive $t$, in the range [0,
360] degrees. For this case ($Z$ chosen as the one marked by the cross sign in Figure~\ref{fig:resmap}), the ranges of these two angles are
$\phi_r\in [90, 107]$ degrees, and $\phi_t\in [118, 128]$ degrees
from the STB point of view toward the Sun (see
Figure~\ref{fig:3DSTB}), with mean values 96 and 123 degrees,
respectively. The supplementary angle to  $\phi_t$ is in the range
of [52, 62] degrees. Correspondingly, the ranges of major radius
and poloidal flux are, $r_0\in [0.81, 1.72]$ AU, and $\Phi_p\in
[1.8, 6.1]$ TWb, respectively.
 \begin{table}
 \caption{The ranges of parameter values for the three different choices of rotation axis orientations. }\label{tbl:2}
 \begin{tabular}{ccccc}
 \hline
 %\multicolumn{2}{c}{<>}
 $Z (r_{sc},t,n)$ & $r_0$ (AU) & $\Phi_p$ (TWb/radian) &
$\phi_r$ (degrees) & $\phi_t$ (degrees) \\
\hline
(0.8663,   0.2722,   0.4188) & [0.77,1.27] & [2.3,4.8] & [101,105] & [118,125]  \\
(0.5239   0.6626   0.5353) & [0.81,1.72] & [1.8,6.1] & [90,107] & [118,128]  \\
(-0.1342   0.9014   0.4116) & [0.75,1.89] & [2.9,7.7] & [77,97] &
[112,115]  \\\hline
%- & - & 11.9 &  & $\sim$129 \\
 \end{tabular}
 \end{table}

The importance of the reconstruction of morphology of flux rope
CMEs from the Sun to the spacecraft intersection by \citet{2010ApJ...715.1524W} is that it
provides guidance in reducing the uncertainty in determining the
flux rope geometry when searching for the set of optimal
parameters in the GS reconstruction. For instance, in this event,
from \citet{2010ApJ...715.1524W}, the spacecraft crossed the flux
rope above the center from the STB point of view toward the Sun
(see Figure~\ref{fig:3DSTB}). Our analysis yielded the magnetic
field configuration consistent with this above-the-center crossing
for the set of optimal rotation axis $Z$ and the resulting major
radius indicated above. Other choices of parameters, such as the
alternative $Z$ axis orientations marked by the circle and the square in
Figure~\ref{fig:resmap} failed to yield such a configuration. In addition, the ranges of angles $\phi_r$ and $\phi_t$ are also more consistent with \citet{2010ApJ...715.1524W} for the chosen optimal $Z$ axis in the middle of the residue contour in Figure~\ref{fig:resmap}.  For
completeness, the corresponding sets of parameters especially in
terms of their ranges are summarized in Table~\ref{tbl:2} for the
three choices of $Z$ axis orientations marked in
Figure~\ref{fig:resmap} in the order from the circle, cross to square symbols for rows 1-3.

A comparison  with the corresponding GS
reconstruction result from \citet{2009ApJM} with the
straight-cylinder geometry shows certain similarities. Both flux
ropes are right-handed (see Figure~\ref{fig:Ptmap}). The axial
field strength is close, 15.4 nT in \citet{2009ApJM} and 14 nT from our analysis.
The inclination of the axis of the flux rope with respect to the
$(r_{sc},t)$ plane is similar, $\sim$ 51 degrees, comparable with our
result, the mean value 57 degrees, for the axis lying nearly  on the $(t,n)$ plane
\citep{2009ApJM}.  The axial (toroidal) and poloidal flux from
\citet{2009ApJM} are 7.2 TWb and 11.9 TWb/AU, respectively,
noticeably  larger than what we obtain in this analysis, 2.7 TWb and 8.2 TWb/radian, correspondingly,
 due to a smaller-size interval used in the current
analysis and relatively large impact parameters in both analyses.

The comparison of the inclination of the flux rope axis with
respect to the $(r_{sc},t)$ plane with \citet{2010ApJ...715.1524W}
shows a deviation of about 20-30 degrees. Our interpretation is that
the GS reconstruction focuses on the magnetic field configuration
embedded within what Wood et al. reconstructed, since the latter
approach based on white-light images yields a density structure
in terms of outer shells bounding the CME, enclosing the GS
reconstruction interval in this case (see Figure~\ref{fig:data}).
Therefore the magnetic flux rope from the GS reconstruction is
largely enclosed by the dense, outer boundary of the CME flux
rope, and they don't necessary share exactly the same apparent
axial orientation although they should not differ dramatically,
say, by 90 degrees, either.

%\subsection{}

\section{Conclusions and Discussion}\label{sec:con}
In conclusion, we have completed the development of the toroidal
GS reconstruction method and conducted the first applications to
both a benchmark study and a real MC event observed by STB. We provide an additional tool in modeling the magnetic flux rope configuration from \textit{in situ} spacecraft data. The
tool is ready for applications to additional events, using
\textit{in situ} spacecraft measurements and reasonable computing
resources. The whole procedure can be completed on a desktop
computer in Matlab within 2-3 hours. The benchmark study given
here and in Paper I indicates that there is significant amount of
uncertainty in recovering the exact geometry of the flux rope. In
addition to deviations in the $Z$ axis orientation and the major
radius (Paper I), the impact parameter, $d_0$, obtained from the
final GS reconstruction of the cross section can differ most
significantly from the exact value, 0.080~AU against 0.015~AU, in
this study. Nonetheless, other physical properties, especially the
magnetic flux, both the toroidal and poloidal component, are
accurately recovered (see Table~\ref{tbl:bench}). We also apply
the method to a real MC event measured by STB that was studied
earlier by both \textit{in situ} modeling by the straight-cylinder
GS reconstruction and the white-light imaging reconstruction.
The GS reconstruction result with the toroidal configuration is
obtained with significant uncertainty. For the optimal rotation
axis, $Z=(0.5239,   0.6626,   0.5353)$, chosen in the
$(r_{sc},t,n)$ coordinate system, the major radius is in the range
$[0.81,1.72]$~AU. We compare the apparent geometry of the flux
rope  configuration with the flux rope volume resulting from
white-light imaging reconstruction at the STB location. We find
that the inclination angles of the central axis of the flux rope
from the two results are generally consistent in terms of the
ranges of the cone and clock angles of the axial direction. But
the clock angle differs by about 20-30 degrees in the plane of the
sky.

It is still not conclusive whether the toroidal geometry is
superior to the original straight-cylinder geometry, despite the
much evolved procedures in the toroidal GS reconstruction. 
The central axis orientation is similar at the point
of spacecraft traversal. The magnetic flux content is very
different, partially due to different time intervals used in these
two separate analysis. For larger major radius, the two
reconstruction results should converge. In other words, as the
major radius $r_0$ goes to infinity, the curvature of the toroidal
flux rope tends to zero, falling back to the original
straight-cylinder analysis. This is somewhat hinted by the open
contours in Figure~\ref{fig:chi2STB}, implying the possible
consistency with a straight-cylinder geometry, for this case. The
added feature in the current analysis, not yet available in the
original straight-cylinder GS reconstruction is the uncertainty
estimates in various quantities, based on proper $\chi^2$
statistics and the measurement uncertainties in magnetic field
components. These are useful, but the caveat is that the results
probably depend on a reliable estimate of the measurement
uncertainties, which is not always available. Nonetheless, the
unique scientific merit of the current fully developed approach
lies in the ability to provide additional characterization of
the magnetic  flux rope variability in the space plasma
environment.

%%%%%%%%%%%%%%%%%%%%%%%%%%%%%%%%%%%%%%%%%%%%%%%%%%%%%%%%%%%%%%%%%%%%%%%%%%%
%% Acknowledgements
%
 \begin{acks}
QH acknowledges partial support from  NASA grants NNX14AF41G,
NNX12AH50G, and NRL contract N00173-14-1-G006. All authors were supported as part of 
 an FST team on flux ropes led by Dr.~Linton and funded by
NASA LWS  award NNH14AX61I
under ROSES NNH13ZDA001N.  \\

\noindent\textbf{Disclosure of Potential Conflicts of Interest}
The authors declare that they have no conflicts of interest.

 \end{acks}

 \bibliographystyle{spr-mp-sola}
 \bibliography{ref_master3}

 \end{article}
\end{document}